\documentclass{article}
\usepackage{spconf,amsmath,graphicx}
\usepackage{tipa}
\usepackage{hyperref}
\usepackage{lineno}
\usepackage{tipa}
\usepackage{caption}
\usepackage{subcaption}
\usepackage{tabulary}
\usepackage{tabularx}
\usepackage{textcomp}
\usepackage{multirow}
\usepackage{multicol}
\usepackage{makecell}
\usepackage{booktabs}

\usepackage{xcolor}

\title{Fundamental Frequency Feature Normalization and \\ Data Augmentation for Child Speech Recognition }
\name{Gary Yeung, Ruchao Fan, and Abeer Alwan\thanks{This work was supported in part by National Science Foundation (NSF) Grant \#1734380.}}
\address{Dept. of Electrical and Computer Engineering, University of California, Los Angeles, USA}

\begin{document}
%
\maketitle
\begin{abstract}

Automatic speech recognition (ASR) systems for young children are needed due to the importance of age-appropriate educational technology.
Because of the lack of publicly available young child speech data, feature extraction strategies such as feature normalization and data augmentation must be considered to successfully train child ASR systems.
This study proposes a novel technique for child ASR using both feature normalization and data augmentation methods based on the relationship between formants and fundamental frequency ($f_o$).
Both the $f_o$ feature normalization and data augmentation techniques are implemented as a frequency shift in the Mel domain.
These techniques are evaluated on a child read speech ASR task.
Child ASR systems are trained by adapting a BLSTM-based acoustic model trained on adult speech.
Using both $f_o$ normalization and data augmentation results in a relative word error rate (WER) improvement of 19.3\% over the baseline when tested on the OGI Kids' Speech Corpus, and the resulting child ASR system achieves the best WER currently reported on this corpus.

\end{abstract}
%

\section{Introduction}
\label{sec:Introduction}

The development of effective child automatic speech recognition (ASR) systems has become important in recent years.
For example, the advancement of child ASR can facilitate the development of teaching and assessment tools for children in educational settings \cite{Tepperman2006, Bunnell2000, Yeung2017} using interactive systems such as social robots \cite{Spaulding2018, Yeung2019, Kennedy2017}.
This is especially relevant for kindergarten-aged children who are just learning to read, write, or type and rely on speech to interact with technology.
Yet, ASR systems for young children still perform quite poorly when compared to adult ASR \cite{Kennedy2017, Yeung2018}.

One of the major hurdles facing the development of effective child ASR systems is the lack of publicly available young child speech databases.
This is especially a concern in an era where deep learning, which requires many hours of training data, is rapidly becoming the primary method of developing ASR systems, using data intensive acoustic models such as bidirectional long-short term memory (BLSTM) networks.
Hence, many child ASR systems complement young child speech with older child speech or even adult speech for training data.
However, there is a large acoustic mismatch between child and adult speech, further complicated by the fact that children's speech acoustics change quite dramatically as they grow \cite{Lee1999, Vorperian2007}.
These changes include formants and fundamental frequency ($f_o$) \cite{Lee1999, Vorperian2007}, two defining features of the speech signal, especially for vowels.
Furthermore, as the age difference between training and testing speakers grows, ASR performance degrades rapidly \cite{Yeung2018}.

One common strategy to account for the acoustic mismatch between speakers is frequency normalization.
This approach attempts to warp the frequency spectra of an utterance given a normalization factor for the utterance and a target acoustic space.
For instance, vocal tract length normalization (VTLN) warps the frequency spectra using a maximum likelihood approach and can be implemented in several ways with varying degrees of success \cite{Stemmer2003, Cui2005, Serizel2014}.
An alternative approach is to use acoustically relevant speech parameters as normalization factors such as subglottal resonances (SGRs) \cite{Guo2015}, the third spectral peak or formant frequency \cite{Cui2006}, and $f_o$ \cite{Yeung2019c}.

Another strategy is to augment the training data by creating additional speech-like features for training data.
There are a number of ways to implement this augmentation such as manipulating the frequency scaling or adding noise \cite{Cui2014, Park2019}.
While data augmentation has not been as readily explored for child speech compared to adult speech, some techniques that have been evaluated include adding noise and reverberation \cite{Wu2019} and using out-of-domain adult data \cite{Fainberg2016}.

In our previous study, we proposed an $f_o$-based normalization technique for child ASR \cite{Yeung2019c}.
In that study, a number of ASR systems were trained using speech from children of various ages and tested using kindergarten-aged children.
While that study demonstrated that $f_o$ normalization was effective for child ASR, the experiments performed assumed that only older child speech was available to train a young child ASR system.
In a more practical situation, we may expect that some amount of in-domain child speech would be available as training data.
Furthermore, we expect to be able to generate additional training data through effective data augmentation techniques.

In this study, we propose a new data augmentation method for training child ASR systems based on the $f_o$ normalization method we proposed in \cite{Yeung2019c}.
This method is capable of generating speech-like features that adhere to the physical properties of speech defined by the relationship between vowel formants and $f_o$.
We show that this data augmentation method is capable of improving child ASR systems adapted from BLSTM acoustic models trained on adult speech.
Additionally, we demonstrate that this data augmentation method can be used alongside $f_o$ normalization for further improvement.

The remainder of the paper is organized as follows.
Section \ref{sec:foNormalization} reviews the $f_o$ normalization technique proposed in \cite{Yeung2019c} and formulates the data augmentation technique.
Section \ref{sec:DatabaseExperiment} describes the databases and experimental setup.
Section \ref{sec:ResultsDiscussion} presents the experimental results.
Section \ref{sec:Conclusion} concludes the paper with a summary and considerations for future work.


\section{Normalization Technique}
\label{sec:foNormalization}

\begin{figure*}[t]
  \centering
  \begin{subfigure}[b]{\columnwidth}
      \includegraphics[width=\columnwidth]{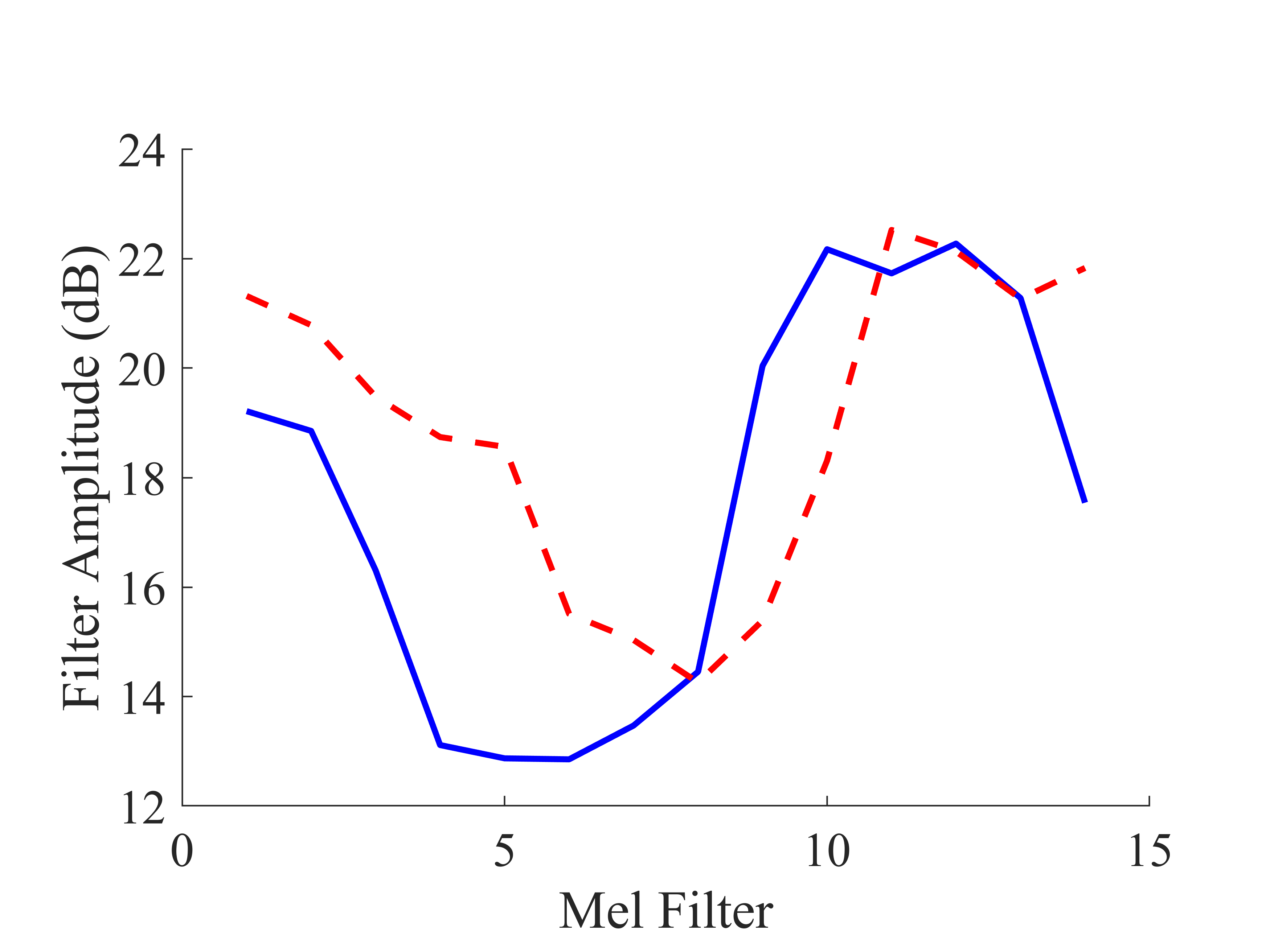}
  \end{subfigure}
  \begin{subfigure}[b]{\columnwidth}
    \includegraphics[width=\columnwidth]{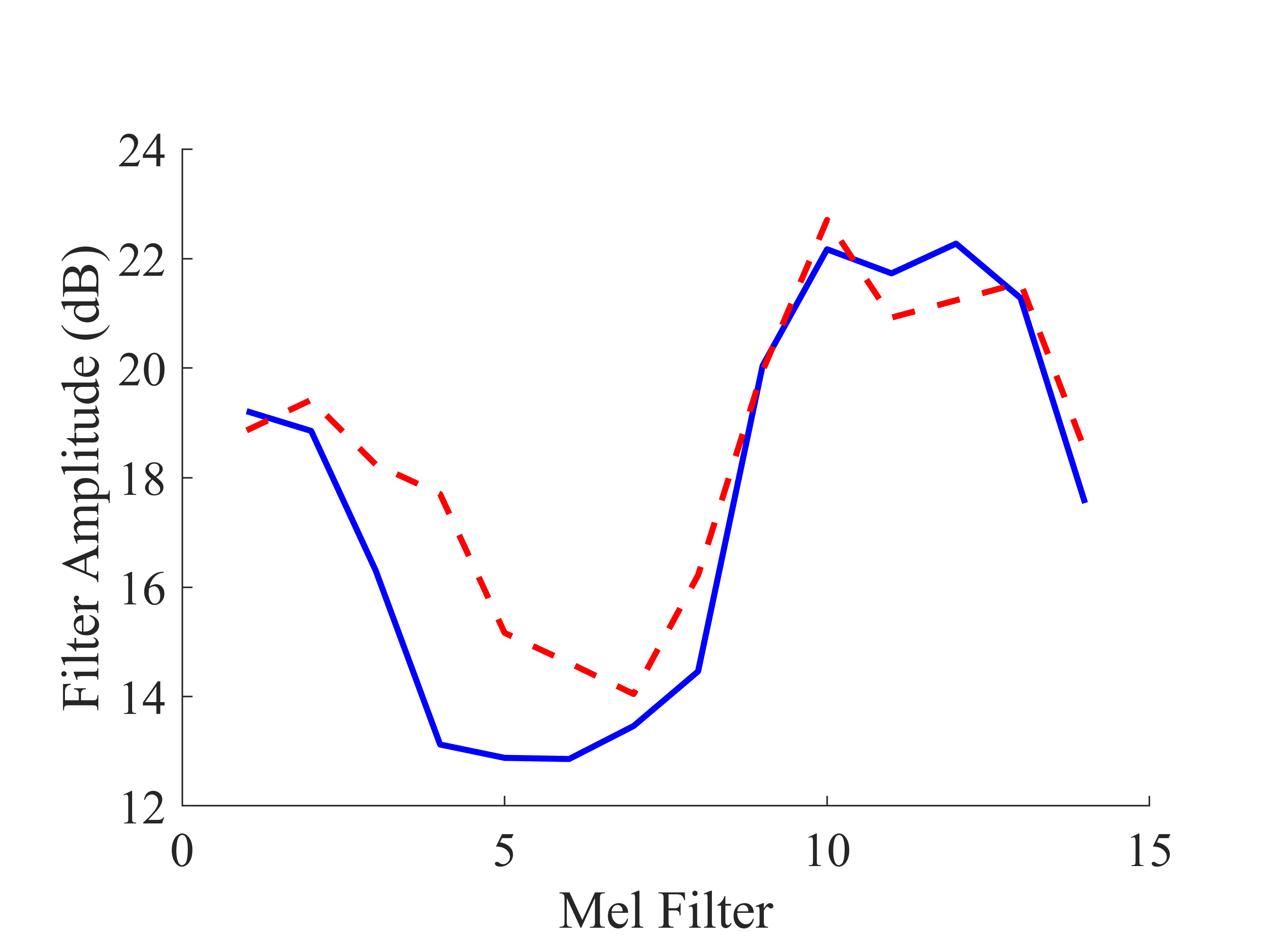}
  \end{subfigure}
  \caption{Mel filter bank outputs of an 18 year old male (solid) and 7 year old male (dashed) saying the vowel /\textipa{i}/, computed with 15 filters and a frequency range of 20 Hz to 6 kHz. The filter outputs are computed both without (left) and with (right) $f_o$ normalization. When normalization is applied, default $f_o$ is chosen to be $f_{o,def}=100$ Hz. The 18 year old male had $f_{o,utt}=106$ Hz, and the 7 year old male had $f_{o,utt}=270$ Hz. The Mel filter bank outputs computed with $f_o$ normalization are much more well-aligned than the filter outputs computed without $f_o$ normalization.}
  \label{fig:CompareF0Norm}
\end{figure*}

\subsection{Relationship Between $f_o$ and Formants}
\label{sec:foFormants}

Several past studies have revealed meaningful connections between $f_o$ and the first several formant frequencies ($F1$, $F2$, $F3$, \ldots) of vowels.
In Bark scale, the tonotopic distances between adjacent formants ($F(x+1)-Fx$ for $x \in \{1, 2, 3, \ldots\}$), along with the tonotopic distance between the first formant and $f_o$ ($F1-f_o$), have proven to be effective at modeling human vowel perception and representing the vowel space \cite{Syrdal1986, Traunmuller1981}.
Furthermore, these tonotopic distances can be equivalently represented by the difference between any formant and $f_o$ ($Fx-f_o$ for $x\in\{1, 2, 3, \ldots\}$).
This suggests that the vowel space can be modeled as a linear relationship (in Bark scale) between formants and $f_o$.
Additionally, studies have found that $f_o$ and formant locations depend on one another in both vowel production \cite{Barreda2013} and vowel perception \cite{Barreda2012}.
A more detailed examination of these relationships can be found in \cite{Yeung2019c}.

\subsection{Normalization Procedure}
\label{sec:NormalizationProcedure}

The frequency normalization technique we proposed for kindergarten ASR in \cite{Yeung2019c} was designed to be applied to any spectral-based feature such as Mel-frequency cepstral coefficients (MFCCs).
The linear relationship between $f_o$ and vowel formants is exploited to formulate a normalization technique using median $f_o$ as the only normalization factor.
As such, the technique attempts to map a speaker's acoustic space, governed by the speaker's $f_o$, to some target space, defined by a predetermined target $f_o$.
Notably, Mel scale was used in \cite{Yeung2019c} instead of Bark scale as the two scales are highly correlated, but either scale can be used in practice.
For consistency, this study will also use the Mel scale.

To perform the $f_o$-based normalization, a default $f_o$, denoted as $f_{o,def}$, must first be chosen.
Before feature computation, $f_o$ extraction is performed across the entire utterance using a reliable $f_o$ detection algorithm.
For this study, we will use the multi-band summary correlogram (MBSC) pitch detection algorithm \cite{Tan2013}.

The median $f_o$ across all voiced frames of the utterance, denoted as $f_{o,utt}$, is then chosen as the normalization factor for the utterance.
The discrete Fourier transform (DFT) of the feature extraction procedure is warped as follows:
\begin{equation}
    \label{eq:foWarp}
    f_{norm} = f_{orig} - (f_{o,utt} - f_{o,def})
\end{equation}
where all frequencies are in Mel scale, $f_{norm}$ is a normalized frequency corresponding to some DFT index, and $f_{orig}$ is the frequency from the original speech spectrum mapped to $f_{norm}$.

An example of this $f_o$ normalization is shown in Figure \ref{fig:CompareF0Norm}.
The Mel filter bank outputs of an 18 year old male and a 7 year old male saying the vowel /\textipa{i}/ are displayed both with and without $f_o$ normalization.
When both utterances are normalized to $f_{o,def}=100$ Hz, the Mel filter bank outputs become more similar.

\subsection{Data Augmentation Procedure}
\label{sec:Regression}

While the $f_o$ normalization procedure attempted to reduce variability between speakers by using Eq. \ref{eq:foWarp}, fixing $f_{o,def}$ to a default value, and adjusting $f_{o,utt}$, an alternative procedure can use Eq. \ref{eq:foWarp} to create variability rather than reduce it.
This can be accomplished by extracting features multiple times from the same speech utterance while adjusting $f_{o,def}$.
By perturbing the feature extraction, we can augment the training data by generating acoustic features that are consistent with the structure of speech defined by the tonotopic distances.
This can be especially useful for training deep neural networks that require a large amount of speech data.
We will refer to this technique as $f_o$ perturbation.
When combined with the $f_o$ normalization procedure, we can simultaneously remove larger inter-speaker variabilities while perturbing the features for additional training data.


\section{Database and Experimental Setup}
\label{sec:DatabaseExperiment}

\subsection{Database}
\label{sec:Database}

Two child speech databases were used in this study.
The first one was the OGI Kids' Speech Corpus \cite{Shobaki2000}.
This corpus contains approximately 100 speakers per American educational grade level, from kindergarten to 10\textsuperscript{th} grade.
Utterances were recorded with a sampling rate of 16 kHz (8 kHz bandwidth).
Both scripted and spontaneous styles of speech were recorded from each speaker.
In this study, we used the sentence utterances from the scripted speech recordings, which consisted of a total of 10,072 sentence utterances from children in grades kindergarten through 5\textsuperscript{th} grade.
Approximately 70\% of the utterances in each grade were used as training data for a total of 7,051 training utterances.
The remaining utterances were used for testing.

The second child speech database was the CMU Kids Corpus \cite{Eskenazi1997}.
This corpus contains 76 speakers between 1\textsuperscript{st} and 3\textsuperscript{rd} grade with two additional speakers from kindergarten or 6\textsuperscript{th} grade for a total of 78 speakers.
Utterances were recorded with a sampling rate of 16 kHz.
A total of 5,180 read sentence utterances were recorded across these speakers.
Exactly 70\% of the utterances in this corpus were used as training data for a total of 3,626 training utterances.
The remaining utterances were used for testing.

For adult data, the LibriSpeech ASR Corpus was used \cite{Panayotov2015}.
For our experiments, we used all the training data from the corpus, which contains 960 hours of adults reading audio books in clean and noisy conditions, for adult model training.

\subsection{Feature Extraction}
\label{sec:FeatureSetup}

The baseline features used for the ASR experiments were 13-dimensional MFCCs with a window size of 25 ms and shift of 10 ms.
These MFCCs were extracted using a 512-point DFT, 23 Mel filters, and bandwidth from 20 Hz to 8 kHz.

The second set of features was similar to the baseline features except the DFT was normalized by the procedure presented in Section \ref{sec:NormalizationProcedure}.
The default $f_o$ was chosen to be $f_{o,def}=100$ Hz representing an adult male $f_o$, and $f_{o,utt}$ was chosen to be median $f_o$ across the utterance, estimated using MBSC pitch detection.
As the value of $f_{o,utt}$ reached as high as 300 Hz, the bandwidth for these features was limited to 6.2 kHz to compensate for the frequency shift upwards.
Notably, this limits the maximum frequency shift of the $f_o$ normalization procedure to approximately 250 Mels.
Both baseline and $f_o$ normalized features were extracted from all utterances.

When applying $f_o$ perturbation, $f_{o,def}$ was adjusted such that the DFT was shifted by $\pm20$, $\pm40$, and $\pm60$ Mels.
Along with the original features, this multiplies the amount of training data by 7.
Data augmentation was applied to the CMU Kids and OGI Kids' training datasets.

Implementation of the feature extraction is relatively simple when using Eq. \ref{eq:foWarp} to compute the DFT frequency shift.
We let $f_{o,def} \in \{$58.52, 72.10, 85.93, 100.00, 114.32, 128.90, 143.74$\}$ Hz where every value in the set is used if performing data augmentation and only 100 Hz is used if performing a standard feature extraction.
Similarly, we let $f_{o,utt} = 100$ Hz when performing a non-normalized feature extraction and choose $f_{o,utt}$ as the median $f_o$ across the utterance when normalizing.

\subsection{Experimental Setup}
\label{sec:ASRSetup}

For both the baseline and $f_o$ normalized features, an adult ASR system was first trained using the LibriSpeech training set.
The acoustic model was a 3-layer BLSTM network with 512 cells in each direction followed by a feed-forward layer that mapped the output of the BLSTMs to senone probabilities.
The input to the BLSTM was 7 frames (3 frames forward and backward in time) to form a 91-dimensional input for the acoustic model. 
The output was approximately 5,700 senone probabilities.
The acoustic model was based on PyKaldi2 \cite{Lu}, while decoding used the Kaldi Speech Recognition Toolkit \cite{Povey2011}.

The adult ASR systems were adapted to child speech using either the CMU Kids or OGI Kids' training data.
The adult ASR system trained with $f_o$ normalization was adapted using child speech features that were extracted with $f_o$ normalization.
Similarly, the adult ASR system trained without normalization was adapted using features that were extracted without normalization. 
Additionally, data augmentation using $f_o$ perturbation was applied on the child training datasets.

A 4-gram language model (LM) trained on Project Gutenburg books was chosen for decoding.
This LM is one of the default language models included in Kaldi's LibriSpeech recipe \cite{Povey2011}.
Adapted ASR systems were evaluated using the corresponding CMU Kids or OGI Kids' testing datasets.
The ASR system trained with $f_o$ normalization was tested using features extracted with $f_o$ normalization.


\begin{table}[t]
    \caption{Word error rates (WERs) of the child ASR experiment using a BLSTM-based acoustic model adapted from adult speech. The left two columns indicate whether $f_o$ normalization (``Norm?'') and data augmentation using $f_o$ perturbation (``Aug?'') were used. WERs for both CMU Kids and OGI Kids' are reported in the latter columns.}
    \label{tab:Results}
    \centering
    \begin{tabular}{ c c | c c }
    \toprule
    \textbf{Norm?} & \textbf{Aug?} & \textbf{CMU Kids} & \textbf{OGI Kids'} \\
    \midrule \midrule
    No  & No  & 16.88 & 6.84 \\
    Yes & No  & 16.93 & 6.50 \\
    No  & Yes & 16.63 & 5.85 \\
    Yes & Yes & \textbf{16.47} & \textbf{5.52} \\
    \bottomrule
    \end{tabular}
\end{table}

\section{Results and Discussion}
\label{sec:ResultsDiscussion}

The results of the child ASR experiments described in Section \ref{sec:DatabaseExperiment} are shown in Table \ref{tab:Results}.
The top row displays the word error rate (WER) of the baseline system (i.e., no normalization or augmentation).
The second row displays the WER of the system using $f_o$ normalization.
The third row displays the system adapted with $f_o$ perturbed child speech features.
Finally, the last row uses both $f_o$ normalization and $f_o$ perturbed adaptation data.
Additionally, we performed the experiment once more using the acoustic model trained only on LibriSpeech.
This system achieved a WER of 37.49\% for CMU Kids and 59.70\% for OGI Kids', significantly worse than any WER in Table \ref{tab:Results}, which demonstrates the importance of child speech adaptation data.

When using adaptation and only applying $f_o$ normalization, the performance of the OGI Kids' system saw a small improvement over the baseline, which has no normalization or data augmentation, but not enough to be significant.
Applying $f_o$ perturbation to the training data, a more substantial improvement was achieved from 6.84\% to 5.85\%.
However, when using the CMU Kids system, no major improvements were observed.

Using both $f_o$ normalization and $f_o$ perturbation resulted in the best performing ASR system for both testing sets.
The OGI Kids' testing set saw a relative WER improvement of 19.3\%, reducing the WER to 5.52\%, and this result is statistically significant at $p<0.001$.
However, the CMU Kids testing set only saw a relative WER improvement of 2.4\%.

While the OGI Kids' testing set saw a larger improvement than the CMU Kids testing set, this result may be expected.
We note that the CMU Kids testing set had a narrower age range (approximately 6-9 years old excluding the two outlier children) compared to the OGI Kids' testing set (approximately 5-11 years old).
The $f_o$ normalization method has been shown to produce larger improvements when the range of ages used in training and testing data is wider \cite{Yeung2019c}.
A similar phenomenon may be occurring for $f_o$ perturbation.
That is, since there is less variability in the CMU Kids testing set, adding additional variability to the training set through $f_o$ perturbation was unnecessary to train the BLSTM.
Meanwhile, with the larger variability of the OGI Kids' dataset, both $f_o$ normalization and $f_o$ perturbation proved helpful.
Furthermore, these techniques may extend to ASR systems using the full OGI Kids' scripted speech dataset, which has been reported to have a WER of 10.8\% \cite{Wu2019}.
Preliminary experiments suggest that $f_o$ perturbation also performs better than VTLN-based data augmentation.


\section{Conclusion}
\label{sec:Conclusion}

This study proposes a new data augmentation method for training child ASR systems based on the $f_o$ normalization method proposed in \cite{Yeung2019c}.
Both normalization and data augmentation methods adhere to the physical properties of speech defined by the relationship between vowel formants and $f_o$.
The two methods can be formulated as a simple shift of DFT bin frequencies in the Mel domain and are implemented by manipulating $f_{o,def}$ and $f_{o,utt}$ in Eq. \ref{eq:foWarp}.
Child ASR systems were trained using these methods by adapting from a BLSTM acoustic model trained on adult speech.
When using both $f_o$ normalization and $f_o$ perturbation, a 19.3\% relative improvement was observed on the OGI Kids' Speech Corpus.
However, a less substantial improvement was observed on the CMU Kids Corpus.
This suggests, that both $f_o$ normalization and $f_o$ perturbation are more effective when the age range of the speakers is large.

There are a number of possible directions for future work.
We plan to evaluate these methods on ASR systems for both children and adults for a situation with increased age and $f_o$ variability between speakers.
We also plan to apply these systems to other child ASR applications such as conversational speech, educational applications, and clinical applications.


\bibliographystyle{IEEEbib}

\bibliography{strings}

\end{document}